\documentclass[%
 reprint,
 amsmath,amssymb,
 aps,
]{revtex4-2}

\usepackage{graphicx}% Include figure files
\usepackage{dcolumn}% Align table columns on decimal point
\usepackage{bm}% bold math
\begin{document}

\preprint{APS/123-QED}

\title{LEO small satellite QKD downlink performance: QuantSat-PT case study}% Force line breaks with \\

\author{V. Galetsky}
\email{vladlen.galetsky@tecnico.ulisboa.pt}
\author{M. Niehus}%
 \email{mniehus@deetc.isel.ipl.pt}
\affiliation{%
 Department of Physics,  Instituto Superior Tecnico - IST, University of Lisbon, Portugal\\
Institute of Telecommunications,  Instituto Superior Tecnico Torre Norte - Piso 10. Av. Rovisco Pais, 1 1049 - 001 Lisbon,  Portugal\\
Departamento de Física, ISEL - Instituto Superior de Engenharia de Lisboa, Rua Conselheiro Emídio Navarro 1, 1959-007 Lisboa, Portugal\\
}%

\date{\today}% It is always \today, today,

\begin{abstract}
In this work, we model and simulate the performance of a quantum key distribution (QKD) downlink from a low earth orbit (LEO) small satellite to an optical ground station (OGS), as integral part of the concept and preliminary design phase of the QuantSat-PT mission. By modelling and simulating in detail downlink transmission channel effects, with emphasis on turbulent and atmospheric losses for BB84 and E91 protocols, we find a consistent set of values for the performance envelope that resolves ambiguities of mission experimental data that had been reported previous to this work.  We obtain for the 4-state BB84 protocol a sifted key rate and Quantum Bit Error Rate (QBER) of 32.1 kbit/s and $4\%$, respectively, for zenith at 750 km orbit. For the E91 protocol the Clauser, Horne, Shimony and Holt (CHSH) test was performed resulting in a correlation factor of $S \in[-2.63\pm0.02,-1.91\pm0.03]$ for the mission. The consistency of these results with the state of the art simulators and its relevance on experimental satellite based QKD is discussed.
\end{abstract}

\maketitle

%\tableofcontents

\section{Introduction} 
\label{Introduction}

Modern cryptography plays an essential role for the security of the transmission of information. Encryption, authentication
and signature scheme processes favor users to stay protected from classical attackers. Most of these protocols are based on RSA \cite{a1,a2} method, where the pillar of encryption is primarily centered on factorization, the discrete logarithmic and elliptic-curve discrete logarithmic problems.

\medskip\noindent
 Recent advances in quantum computing \cite{a3} performed by Google quantum AI claimed to reach for the first time quantum supremacy in 2019 \cite{a4}. More recently in 2021 by studying a new series of $\textit{Quantum Falcon}$ processors, IBM increased the quantum volume of circuits to 64 for their current architecture \cite{a6}. Moreover in the same year, it has been claimed that the superconducting quantum processor, $\textit{Zuchongzhi}$, also achieved quantum supremacy \cite{a5} creating an opening for a new era of quantum computing. However, to this day, no architecture reaches more than 127 qubits. Assuming that in the next years a powerful quantum computer reaches an order of twenty billion qubits, by running Shor's or Grover's algorithm \cite{a7} it will lead to a collapse of most modern classical public key cryptosystems. Despite of existing symmetric key algorithms that are resistant to quantum computation attacks (AES-256), not all of them share this trait. One of the solutions, Quantum key distribution (QKD), is assessed in this article, in low earth small satellite QKD downlink configuration, as a pathway for privacy of communication.

\medskip\noindent
Due to the required trade-off between key rates, QBER and distance for a terrestrial optical fiber based QKD link, a satellite solution study turns into a necessity to assess the advances of quantum computers. Moreover, with the advances in the telecommunications sector future optical and quantum networks are being proposed as a viable solution to support future 6G technologies \cite{Nguyen_2021}, ergo, leaving a prerequisite to test the feasibility and sustainability of such satellite missions.

\medskip\noindent
With this in mind, in this article a feasibility analysis is presented for the first Portuguese QKD satellite mission QuantSat-PT, as integral part of the concept and preliminary design phase. The mission aims to perform and test on a 3U Cubesat a quantum key exchange downlink to an optical ground station (OGS) localized in a rural region (Alqueva) on the Portuguese mainland. The mission is based on a 4-state BB84 protocol, with possible expansion to entangled state E91 protocol type downlink which is also analysed here by quantifying Bell test results in the case study. 

\medskip\noindent
In this article we analyze the QuantSat-PT mission, by modeling and simulating time dependent sifted key rates and QBER for the BB84 protocol. An analysis is also performed for the E91 protocol via a time dependent CHSH type Bell test during the satellite pass-by. The work is structured as follows: In Section 2, we contextualize and present an overview of state-of-the-art QKD space missions and simulators. In Section 3, we describe the methodology used for orbit selection and protocol simulation. In Section 4, the analysis focuses on the transmission channel losses by including natural and artificial background, atmospheric and turbulent behaviour as well as the off-pointing behaviour of the S/C. In Section 5, we discuss the results and expected performance, both for the BB84 and E91 protocols, comparing the results with the state of the art. In Section 6, we conclude by giving an outline of the main results.
\begin{figure}
\label{Method} 
\centering
\includegraphics[width=0.48\textwidth,clip]{ 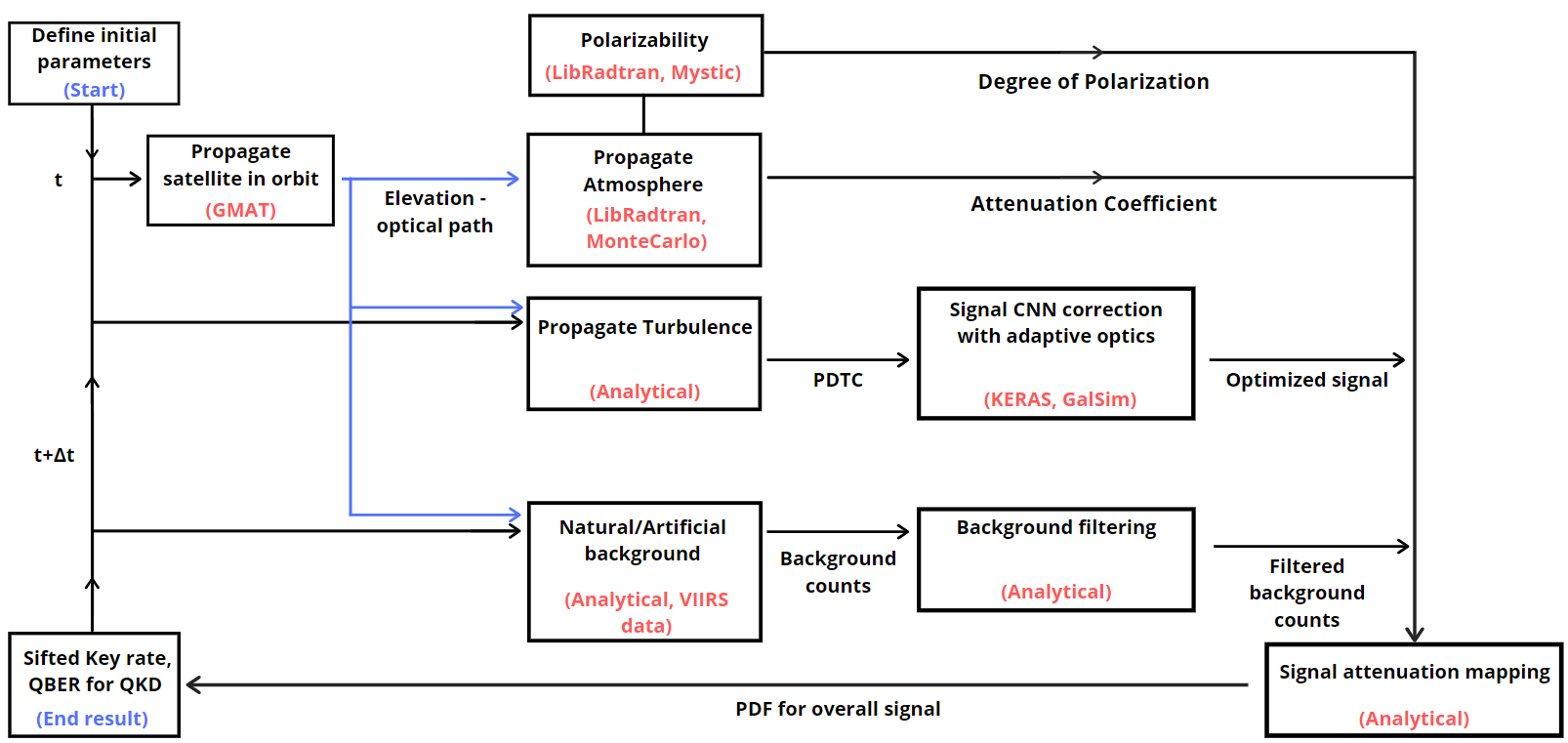}
\caption{
Architecture of the simulator for the BB84 4-state protocol. Depolarization, geometric, atmospheric, background and turbulence channel losses are considered for each satellite time-step in order to obtain the sifted key rate and QBER of the S/C. In red, the used methodology or adapted software is presented for each section.}
\end{figure}

\section{Overview}
\label{Overview}

\subsection{QKD satellite missions}
\label{QKD satellite missions}

\medskip\noindent
To benchmark the mission we shall briefly review relevant state of the art satellite based QKD missions.

\medskip\noindent
Micius was successfully
launched on August 2016 \cite{v1} from Jiuquan, China, and now orbits at an altitude of about 500 km in
LEO. One of the satellite payloads had a BB84 decoy-state QKD transmitter at a wavelength of 850
nm cooperating with Xinglong ground observatory station. At optimal distance of 600 km, it achieved
a QBER $\sim$ 2$\%$ and a sifted key of 14 kbps. In 2017 the mission concept went into a second phase
\cite{v1}, establishing a space-to-ground two-downlink channel creating a 1200 km distance QKD between
two ground-stations (GS), however, a secured node configuration was assumed,
meaning that no eavesdropping was considered (QBER $\sim$ 8.1$\%$ with sifted key rate of 1.1 Hz). The measured overall
two-downlink channel attenuation was at peak 82 dB which was higher than predicted.

\medskip\noindent
On the other hand, SOCRATES \cite{T2} micro-satellite was first launched in 2014, with the main objective of
technology demonstration for position and attitude control. SOTA, lasercom payload, had a secondary
mission in 2016 of creating a B92-like QKD protocol at 800-nm band to perform the first-time quantum limited demonstration from space. For the B92 protocol an emulated QBER $\in[3,6]\%$ was obtained for the mission active time. From the results given \cite{v2}, it’s mentioned that the total loss budgets from the simulation analysis and the real data losses received were off from a range of 29.5 dB to 13.8 dB. Thus, pointing to unmet
simulation conditions for QKD analysis due to the complexity of the problem in study. In the work of \cite{v2},
these values are attributed to atmospheric scintillation, which typically could change the losses by an 
order of magnitude. From these results, the discrepancy between simulations and experimental data fully motivates to have a more detailed study on satellite QKD communications.

\medskip\noindent
\subsection{QKD space simulators}
\label{QKD space simulators}

To understand the motivation for the approach taken for this article a diverse number of state of the art simulators are presented. Mainly from the works of \cite{v4}, QUARC \cite{v5}, \cite{v6} and some statistical methods from \cite{v7,v8}. 
\medskip\noindent

Burgoin et al. simulated and calculated the expected performance for a year-long 600 km satellite conducting a QKD link at 670 nm for a sun-synchronous orbit implementing a decoy state BB84 protocol. A Rayleigh–Sommerfeld diffraction \cite{v10} was considered with a custom beam profile with a convoluted pointing error. MODTRAN \cite{v9} was used for the atmospheric attenuation calculation, artificial and natural background was considered. An in-depth comparison between up-link and downlink approaches were also performed. For a 600 km downlink, considering a beam waist of $w_{0}=0.05$ m and diameter of receiver of $D_{R} = 1.0$ m at the GS, the author's achieve a mean QBER of $4.3\%$ between zenith angles of $\theta_{zen}\in[0,70]$ deg. For high zenith angles it reaches up to $11\%$.

\medskip\noindent
QUARC opted to simulate how a constellation comprising 15 low-cost 6U CubeSats with the BB84 protocol usage can form a secure communication backbone for ground-based and metropolitan networks across 43 GS in United Kingdom. It has an in-depth study of the satellite tracking and the telescope FoV. Something of the utter importance for high precision laser communications to guarantee an optimal key rate. For a 574 km orbit, with a laser wavelength of 808 nm and assuming transmitter and
receiver apertures of 0.090 m and 0.7 m respectively, the author's obtained a loss range from approximately 47 dB to 35 dB during the satellite's active time.

\medskip\noindent
A different take on the approach looks into the Probability Distribution of the Transmission Coefficient (PDTC), a statistical interpretation for the off-pointing, turbulence disturbance and atmospheric effects. In the work of \cite{v7} the
effect of channel fluctuations is studied in continuous variable QKD, using in the simulator a derivation of the equations for the secret key rate over generic fading channels. Overall, for a 800 km orbit a downlink transmissivity of $1.8\%$ of the total sent signal is achieved.

\medskip\noindent
Another method, from the work of \cite{v6} considers that imperfections from the truncation of the border in optical elements, conditions in the far field an additional broadening of the beam. Thus, it considers an imperfect quasi-Gaussian beam in a turbulent environment characterized by a probability distribution function (PDF). The author's work achieves for Cubesats a $QBER=3\%$ at low zenith angles going up to $QBER=14\%$ at zenith angles above $\theta_{zen}=75\%$. The results on different approaches come close to the same order of magnitude.

%PIL

\section{Methodology}
\label{methodology}

\medskip\noindent
For this article the methodology towards a realistic simulation within mission constraints for 4-state BB84 protocol is presented in Fig \ref{Method}.

\medskip\noindent
The objective of the simulator is to validate the feasibility of the mission and to adapt the parameters according to mission requirements. Furthermore, as seen in Fig.\ref{Method} we can quantify and analyse the feasibility by obtaining the QBER and sifted key rate for the BB84 protocol, and by performing the CHSH test and validating the correlation factor (S) for the E91 protocol. The simulator is divided in different sections propagating in parallel at each time step. 

\medskip\noindent
As seen in Fig \ref{Method}, the simulator has a bottom up structure, where by defining specific channel losses we increase the robustness and complexity of the simulator. An interative behaviour is also present, where by defining the initial parameters of the ground station, satellite optical apparatus, S/C orbit and atmospheric conditions we can infer on the next state of the system. 

\medskip\noindent
In Fig \ref{Method}, from left to right, in order to define the initial parameters of the system, we would first need to determine the S/C orbit for which the mission requirements remain valid.

\medskip\noindent
After the S/C orbit definition, a geometric channel loss is introduced as a toy model by propagating a Gaussian beam from the S/C towards the GS. This allows us to ensure that the necessary requirements for the optical apparatus and the ground station are also met by studying the QBER and key sifted key rate for such a model. 

\medskip\noindent
After optimizing the initial parameters accordingly to the mission requirements, as seen in Fig \ref{Method}, four additional channel losses were modelled and introduced:

\begin{itemize}
  \item A natural and artificial background profile and noise ratio.
  \item A static atmospheric profile which provides via absorption and scattering the signal's attenuation.
  \item The photon depolarization losses from Rayleigh scattering.
  \item A dynamic atmospheric turbulent profile, by studying in depth the scintillation, atmospheric beam spreading and beam wandering effects on our signal by presenting themselves as channel losses.
  
\end{itemize}

\medskip\noindent
Later on, the off-pointing behaviour of the S/C was also introduced as an additional channel loss by providing with a statistical model based on the PDTC. Hence, allowing to statistically map the signal losses and as a result allowing to create more robust scenario for the sifted key rates and QBER for the 4-state BB84 protocol.  

\subsection{Orbit selection}
\label{Orbit selection}

\medskip\noindent
At first we start by propagating the 3U Cubesat using the GMAT software \cite{zed10}. A low earth orbit (LEO) of 400, 500, 600 and 750 km was considered, propagated during 7 days. In Tables \ref{parameter} and \ref{parameter2} the main parameters for the optical link and the orbit simulation are shown, respectively. As seen, an inclination of 98.00 degrees was selected due to a requirement to perform the quantum key exchange at least once during the night cycle, after 23:00 (GMT+1). This requirement accounts for the lowest natural background environment.

\medskip\noindent
The propagator errors were also calculated by comparing the Runge-Kutta 89 (RK89) \cite{Kutta} numerical integrator with the RK-DP853 ($\pm10^{-3}$ km), RK-DP78 ($\pm10^{-5}$ km), RK-DP45 ($\pm10^{-4}$ km) and RK4 ($\pm10^{-1}$ km), for the position and velocity vectors. An end of life orbit analysis was also performed by using the OSCAR ESA's Drama tool \cite{zed11}. Drama provides an endorsing accordingly to the space debris mitigation guidelines for the end of life cycle of an orbit. OSCAR allows to use Monte-Carlo with generated data of solar flux and geomagnetic activity on the satellite’s trajectory, deviating it.

\begin{figure}
\centering
\includegraphics[width=0.48\textwidth,clip]{ 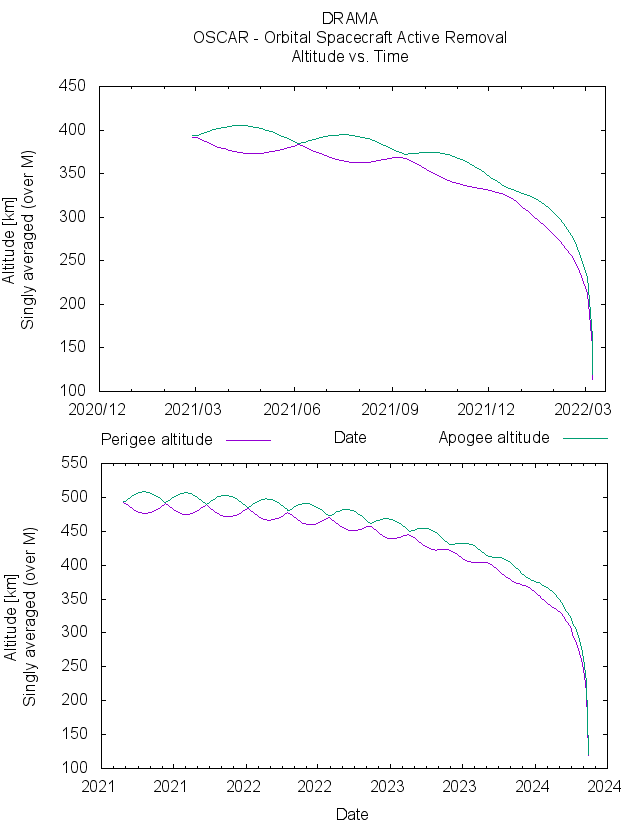}
\caption{Perigee and Apogee calculation for the end of life orbit for 400 km and 500 km, respectively }
\label{DRAMA}       
\end{figure}

 To ensure the mission life cycle requirement of at least 3 years for the 500, 600 and 750 km orbits, Drama was used to identify such limit as can be seen between the 400 km and 500 km orbits in Fig.\ref{DRAMA}.

\begin{figure}
\centering
\includegraphics[width=0.50\textwidth,clip]{ 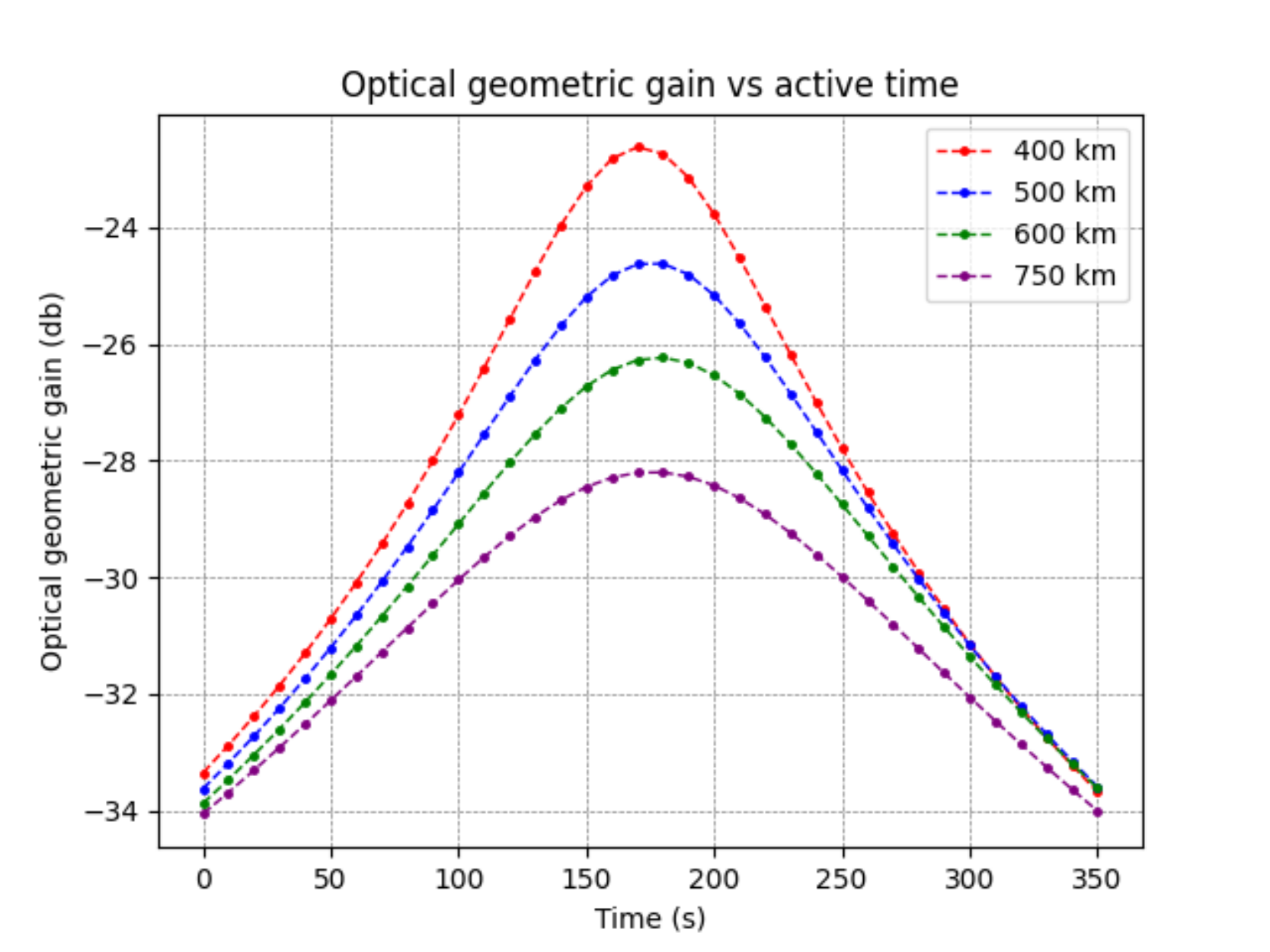}
\caption{Optical gains considering a Gaussian beam between the S/C and GS for the 400 km, 500 km, 600 km and 750 km orbits, respectively. }
\label{key}       
\end{figure}

\medskip\noindent
In order to select an orbit an optical geometric analysis was performed according to the reference \cite{tu2}, as seen in Fig.\ref{key}. Between 400 km and 750 km orbits a 5.1 dB geometrical loss was estimated at zenith by considering in the same conditions a propagation of a Gaussian beam between the S/C and GS. Moreover, the drag coefficient was also taken into consideration for the orbit selection based on the work of \cite{mona1}, were the oscillatory behaviour is greatly attenuated by three orders of magnitude by changing from a 400 km to a 750 km, thus, influencing the off-pointing behaviour of the satellite. By considering all factors from the geometric performance a 750 km orbit was selected for the mission. 

\begin{table*}
\centering
\caption{
The setup of parameters for the optical simulation with a Gaussian beam propagating through the optical path from the S/C to the GS.}
\begin{tabular}{ |p{3.5cm}|p{3cm}|p{1cm}|p{3.5cm}|p{3cm}|p{1cm}| }
\hline
\textbf{Parameter} & \textbf{Value} & \textbf{Unit} & \textbf{Parameter} & \textbf{Value} & \textbf{Unit}\\
\hline
Diameter of Transmitter ($D_{T}$) & $0.03$ & m & Diameter of Receiver ($D_{R}$) & $2.0$ & m\\
\hline
Pulse rate & $100.0$ & MHz & MPN & $0.5$ & - \\
\hline
Optical efficiency ($\eta_{opt}$) & $0.5$ & -  & Wavelength & $850.0\pm1.0$ & nm   \\
\hline
Dark count probability ($Y_{0}$) & $10^{-5}$ &  -   & Basis Misalignment ($e_{det}$) & $0.033$ &  -  \\
\hline
Quantum efficiency ($\eta_{quant}$) & $0.4$ &  - & FoV & $7.14\times10^-4$ &  rad \\
\hline
Total brightness & $2.22\times10^{-4}$ & $cd.m^{-2}$ & Artificial brightness & $5.10\times10^{-5}$ & $cd.m^{-2}$  \\
\hline
GS Latitude & $38.21585$ &  deg & GS Longitude & $-7.58783$ &  deg \\
\hline
\end{tabular}
\label{parameter} 
\end{table*}

\begin{table*}
\centering
\caption{
The setup parameters for the 400 km S/C orbit simulation via GMAT. Similar parameters have been considered for the 500 km, 600 km and 750 km cases, by varying the semi-major axis in the simulation to 6871.00 km, 6971.00 km and 7121.00 km, respectively.}
\begin{tabular}{ |p{3.5cm}|p{3cm}|p{1cm}|p{3.5cm}|p{3cm}|p{1cm}| }
\hline
\textbf{Parameter} & \textbf{Value} & \textbf{Unit} & \textbf{Parameter} & \textbf{Value} & \textbf{Unit}\\
\hline
Drag coefficient & $2.20$ & - & Reflectivety coefficient & $1.30$ & -\\
\hline
GS elevation visibility & $10.00$ & deg & Integrator & RK89 &  - \\
\hline
Eccentricity  & $1.21\times10^{-16}$ & - & Semi-major axis  & $6771.00$ & km  \\
\hline
Inclination  & $98.00$ & deg   & RAAN & $295.00$ & deg  \\
\hline
Argument of Perigee  & $0.00$ & deg & True Anomaly & $1.48\times10^{-6}$ &  - \\
\hline
\end{tabular}
\label{parameter2} 
\end{table*}

\subsection{QKD model definition}
\label{Protocol definition}

\subsubsection{4-state BB84 protocol}
\label{4-state BB84 protocol}

\medskip\noindent
In order to calculate the key rate that we would obtain at GS, we should first estimate the number
of photons sent by the single photon source (SPS) considering a time step of $\Delta t$ = 10 s between each key exchange. Afterwards, the expected number of photons is calculated from the expected power received at GS as seen from the
following equations:

\begin{gather}
P(d,D_{R})=P_{0}\left(1-e^\frac{-D_{R}^2}{2w_{d}^2}\right)
\label{eq:3.3}\\
Q = \eta_{q}\eta_{opt}\Delta t \frac{P(d,D_{R})}{E_{\gamma}}
\label{eq:3.4}
 \end{gather}

\medskip\noindent
$P(d,r_{0})$ is the expected power received at GS. The Gaussian spot size ($w_{d}$) is also provided in terms of optical path ($d$). As stated in Table \ref{parameter}, $\eta_{q}$ and $\eta_{opt}$ are the quantum and optical efficiencies respectively. $Q$ is the expected photon fraction received at GS.

\medskip\noindent
Considering that the standard nomenclature of QKD for Alice and Bob is being used. From Equation \ref{eq:3.4} the sifted key rate ($K$) can be also calculated by comparing the Alice's and Bob's basis ($\delta_{A_{i}}^{B_{i}}$) as seen in Equation \ref{eq:3.6}. An added Gaussian white noise with standard deviation of $\sigma=0.1$ was considered to account for systematic errors caused by thermal and hardware oscillations. The multiplicative behaviour of hardware dependant noise comes from its reliance on the state of the receiver for each time step, hence, influencing in this way on the sifted key rate \cite{gudd}: 
\begin{gather}
K = \frac{\sum_{i}^{n_{t}}\delta_{A_{i}}^{B_{i}}e^{-\frac{x_{i}^{2}}{2\sigma^2}}}{\sigma n_{t}\sqrt{2\pi}}Q(P(d,D_{R}))
\label{eq:3.6}
 \end{gather}

\medskip\noindent
As for the QBER, we have used the following equation to describe it:

\begin{gather}
QBER=\frac{\frac{Y_{0}}{2}+e_{det}(1-e^{-\eta L(\theta)\mu(\theta)})}{\frac{Y_{0}}{2}+1-e^{-\eta L(\theta)\mu(\theta)}}
\label{eq:qber}
 \end{gather}
%-

\medskip\noindent
As stated in Table \ref{parameter}, $Y_{0}$ and $e_{det}$ are the dark count probability and the basis misalignment respectively. L is the optical path which is dependent on elevation ($\theta$), $\eta$ contains the optical and quantum efficiencies. Finally, $\mu(\theta)$ is the generalized loss function which is also geometrically dependent on elevation. It is characterized by the realistic signal attenuation profile, from all the added channel losses in the simulator. 

\subsubsection{E91 protocol}
\label{E91 protocol}

\begin{figure}
\centering
\includegraphics[width=0.45\textwidth,clip]{ 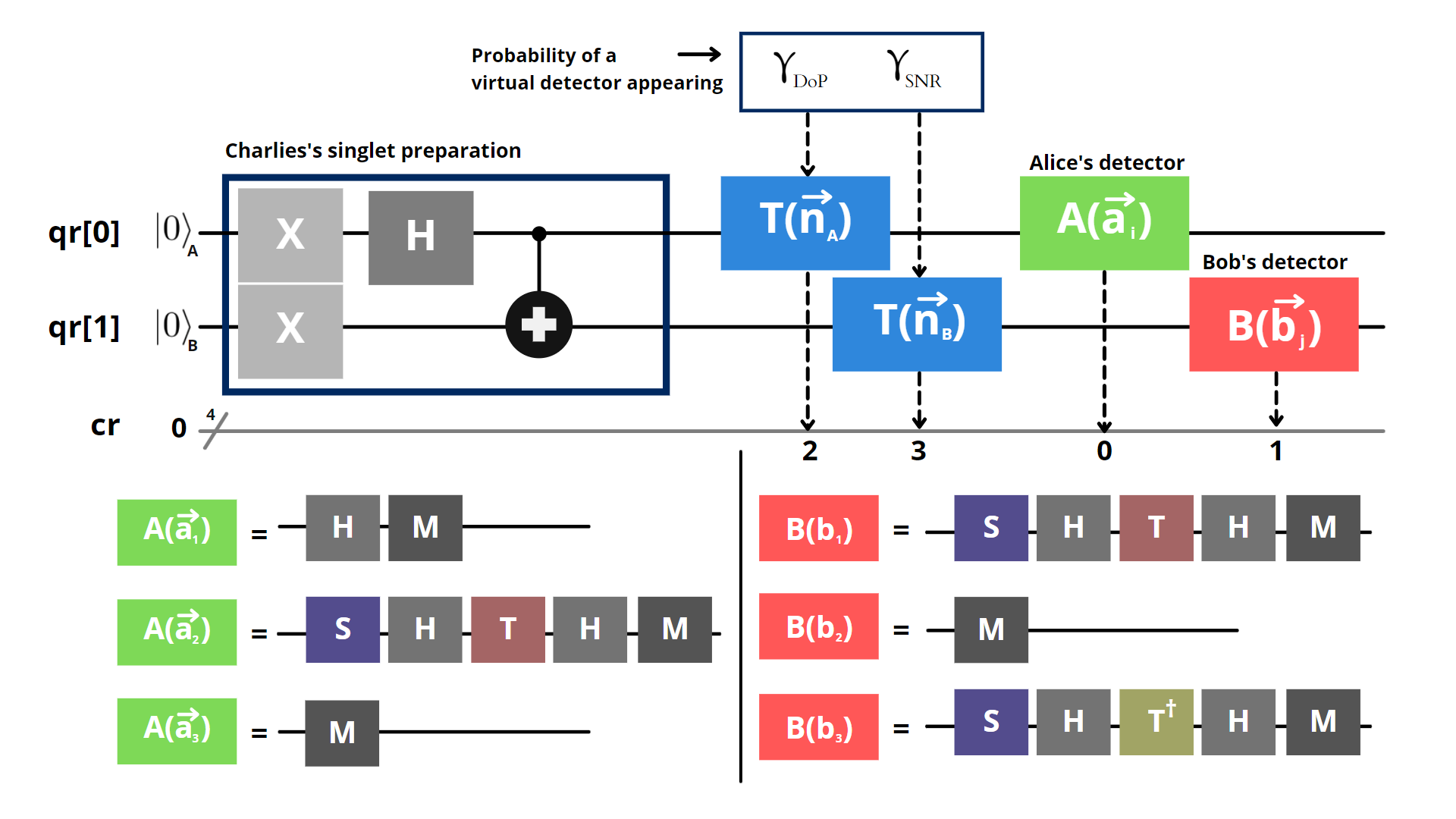}
\caption{Schematic of the quantum circuit used for the E91 protocol. Below, with the use of quantum gates we can rotate the basis for Alice's and Bob's detection.}
\label{E91M}       
\end{figure}

\medskip\noindent
For the 4-state BB84 protocol we have considered an averaged photon number approach to calculate the QBER and sifted key rate. Instead, for the E91 protocol, we shall use the Qiskit Python toolbox 5, to create a modified E91 protocol which considers different attenuation effects as seen in Fig \ref{E91M}.

\medskip\noindent
We measure the feasibility of the mission by performing the CHSH test, by measuring the correlation coefficient (S) between the S/C and GS basis as presented below:

\begin{gather}
S=E(a_1,b_1)-E(a_1,b_3)+E(a_3,b_1)+E(a_3,b_3)
\label{eq:3.3}\\
E(a_i,b_j)=\frac{N_{++}+N_{--}-N_{+-}-N_{-+}}{N_{++}+N_{--}+N_{+-}-N_{-+}}
\label{eq:E91}
 \end{gather}
 
 \medskip\noindent
 Where $N_{xy}$ is the number of coincident counts in a pair of x, y states from the Alice's transmitted and Bob's received photons. Without any environmental or geometrical perturbations considered, if the correlation coefficient follows $|S| \leq 2$, it means that either the
received photons are not truly entangled (which could be
due to an attempt to eavesdrop) or that there is a high magnitude of channel losses \cite{bobi1}.

\medskip\noindent
For our mission, this means that CHSH test cannot identify an eavesdropper if the signal is disturbed to a point where the following inequality $|S| \leq 2$ is always valid.

\medskip\noindent
As seen in Fig \ref{E91M}, we have created a quantum circuit which not only performs the CHSH test, but also adds two virtual detectors which appear in the optical path between the transmitter and receivers, each having a role as photon channel losses. The appearance of those detectors are utterly dependant on the channel losses which correspond to a probability of a photon to shift their basis randomly on its appearance, hence, deviating the correlation factor (S).  

\medskip\noindent
\section{Losses}
\label{Losses}

\subsection{Background modelling}
\label{Losses}

\medskip\noindent
To study natural and artificial background light impact, we have used the VIIRS open source data which provides satellite imagery of Earth in different ranges
of wavelength along time. This analysis was centered on the Alqueva (Portugal) region after 23:00 (GMT+1) as was stated by the requirements to diminish the background number of counts.

\medskip\noindent
For the mission we shall consider the day and night band filter (DNB) which has sensitivity within the visible (V) / infrared range (IR) of $\lambda\in[500,900]$ nm
which contains the operational wavelength for our mission $\lambda = 850\pm1$ nm. To have a change of reference from the Suomi NPP satellite to the GS one, we have used the ATLAS 2015 data-set as seen from the work of \cite{nasa2} to find the total brightness viewed from the GS.

\medskip\noindent
In order to calculate the total number of background counts at zenith, we have used the following equation: $N_{tot}=\frac{1}{E_{\lambda}}\{ (H_{nat}+H_{art})\times \pi (FoV)^2 \times q_{eff} \times \frac{\pi}{4} D_{R}^2\}$ with an added quantum efficiency term $q_{eff}$.
$N_{tot}$ is the background number of counts and $D_{R}$ is the diameter of the telescope. $H_{nat}$ and $H_{art}$ are the natural and artificial night sky brightness respectively. FoV is the field of view of the GS and $E_{\lambda}$ is the energy of each photon with wavelength $\lambda$.

\begin{figure}
\centering
\includegraphics[width=0.48\textwidth,clip]{ 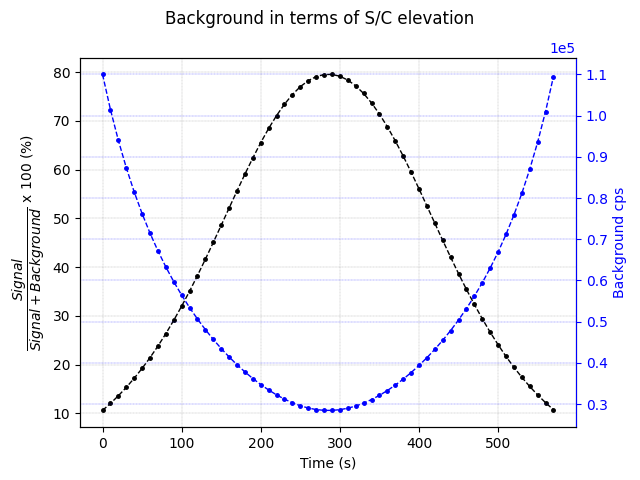}
\caption{
Fraction of signal received at GS during active time. At zenith, $79.8\%$ of the received photons come from the S/C.}
\label{BACK}       
\end{figure}

\medskip\noindent
With the intention of generating a dependence of the number of background counts in terms of the S/C elevation, we follow the work of \cite{nasa3,nasa4, nasa5,net5} which allows us to integrate the Rayleigh airmass density as a function of sky brightness at different satellite elevations. Therefore, a plot is obtained for the total background and the signal ratio in terms of S/C elevation as seen in Fig \ref{BACK}. For the simulator we consider a low lit moon, meaning that background polarization from this source due to Rayleigh scattering is negligible. The time gating filtering is considered in this simulation by decreasing up to an order of magnitude the received background. These results follow the work of \cite{dam3} where the anti-bunching improves up to the same order by using a 2.5 ns time-gating filter.

\medskip\noindent
The background results have been compared with the ESO-Paranal data for IR band achieving similar sky-brightness variations ($\Delta m$) at different S/C zenith angles ($\theta$), with $\Delta m\in[0.02,0.55]$ from $\theta\in[0,60.0]$ deg, respectively.

\medskip\noindent
For our mission, as seen in Fig \ref{BACK} the background is within a range of $B\in[0.3, 1.1] \times10^5$ cps leading to a $S_F=\frac{Signal}{Background + Signal}\times 100 \% \in[10.2,79.8]\%$ depending on the zenith angle of the S/C during active time, $\theta\in[80,0]$ deg, respectively.

%IPL

\medskip\noindent
\subsection{Atmospheric design}
\label{Atmospheric losses}

To construct the atmosphere we shall use an open source software called $\textit{libradtran 2.0.4}$ \cite{bib2}. We have used and modified the Mystic software, which uses Monte-Carlo to solve the polarized radiative transfer equation in 1D \cite{bib8}. Mystic was developed to perform photon forward tracing, where individual photons
are traced from their source to their random paths through the 1D atmosphere, hence being a viable solution for depolarization calculations.

\medskip\noindent
For the atmosphere simulation, we have used the following input parameters such as a mid-latitude summer profile as well as a gamma distribution for cloud droplets. A rural aerosol environment based on the OPAC dataset \cite{bib14} was also considered, the following data provided with the concentration of soil particles, sea salt mixtures and desert dust particles \cite{bib15}. The aerosol visibility was defined accordingly to experimental data in the same rural environment, reaching up to 23.0 km from the GS.
 A REPTRAN \cite{bib10} file was also provided which performs a band parametrization at the top of the atmosphere with a fine bandwidth resolution of 1.0 $cm^{-1}bin$ \cite{bib6}. 

\begin{figure}
\centering
\includegraphics[width=0.48\textwidth,clip]{ 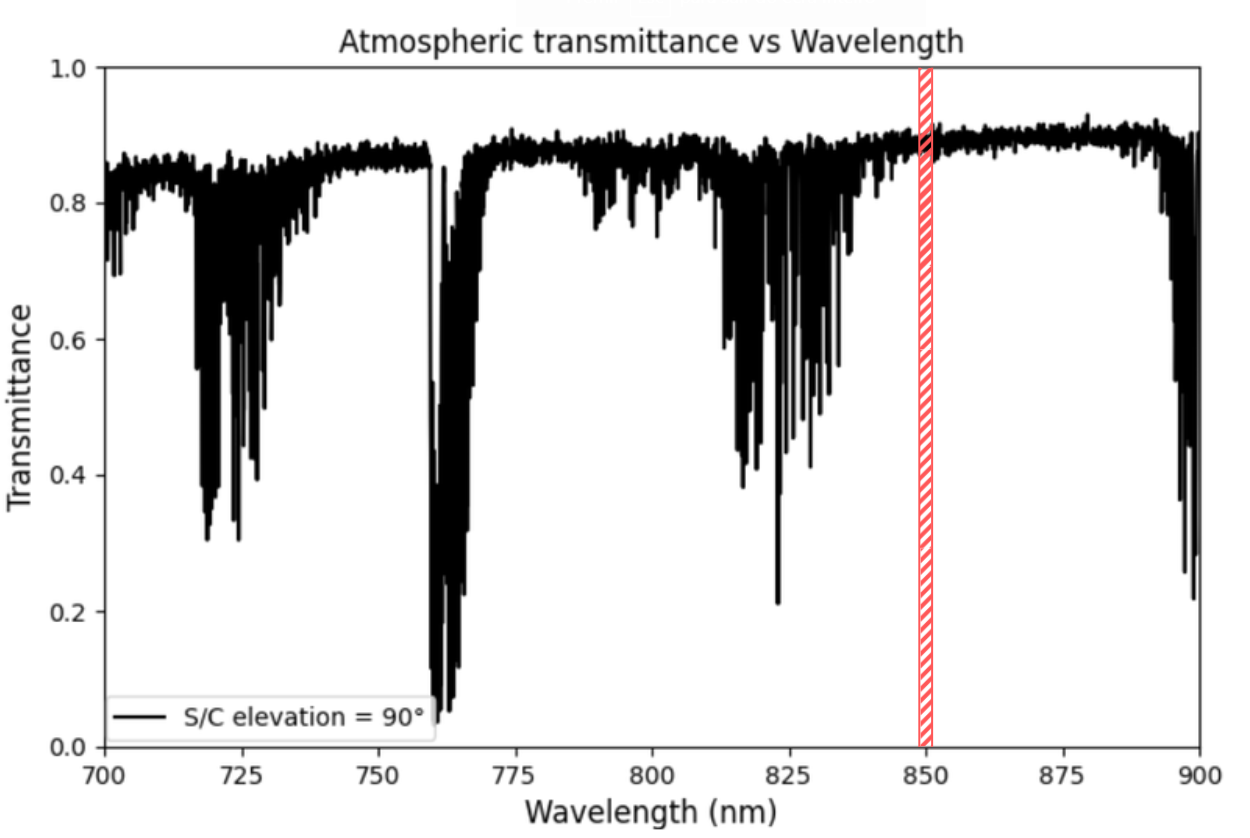}
\caption{S/C photon transmissivity at zenith propagating along the atmosphere at different wavelengths. }
\label{tras}       
\end{figure}

\medskip\noindent
In Fig \ref{tras} we show the obtained results of our simulation where the transmissivity is plotted as a function of different wavelengths. As observed, we obtained a transmissivity of the signal for this atmosphere along zenith of $\tau_{atm}=0.851^{+0.037}_{-0.018}$.

\medskip\noindent
We have also compared our results for the atmospheric transmissivities along the elevation with a theoretical model which considers: $\tau_{atm}=\tau_{zen}^{sec(\theta_{zen})}$ \cite{v7}. Here $\tau_{zen}$ is the optical transmissivity at zenith, and $\theta_{zen}$ the corresponding zenith angle. A similar transmissivity profile was obtained between the two methods along different S/C elevations. We also compared our results with current state of the art methods \cite{v4,tu2} which used the MODTRAN 6 \cite{bib12} software, obtaining plots within the same order of magnitude. 

\medskip\noindent
\subsection{Degree of Depolarization}
\label{Atmospheric losses}
For polarization dependent QKD protocols, there's a signal attenuation occurring from the Rayleigh scattering which persists in different layers of the atmosphere conditioning the photon degree of polarizability.

\medskip\noindent
To classify photon polarization we have used the Stokes parameters ($I$, $Q$, $U$, $V$). For this Monte-Carlo simulation we have considered that all photons sent from the source are linearly polarized $\gamma=(1,1,0,0)$ during a timestep of 10s between acquisitions for each S/C elevation. By defining the degree of polarization as $DoP=\frac{\sqrt{Q^2+U^2+V^2}}{I}$, we can plot it, as well as propagate the mean error for each S/C elevation as seen in Fig.\ref{DoP}:

\begin{figure}
\centering
\includegraphics[width=0.40\textwidth,clip]{ 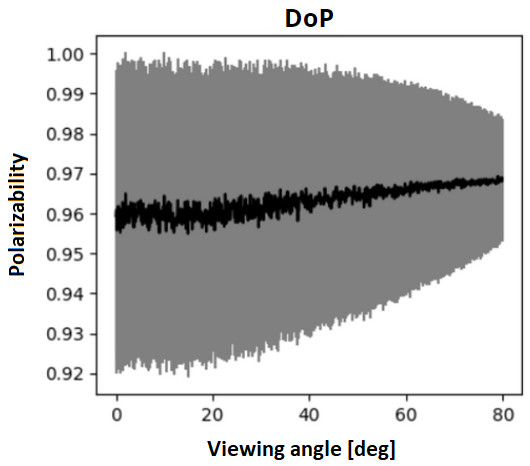}
\caption{Degree of photon depolarization in terms of S/C elevation.}
\label{DoP}       
\end{figure}

\medskip\noindent
At the horizon, we have obtained $DoP(\%) = 96.1 \pm 3.9 $ $\%$ which means that depolarization can disturb the signal within a range of $[0.2,8.1]\%$. By increasing the elevation, we observe a decrease in the Monte-Carlo root (MSE) reaching a $DoP(\%) = 96.8\pm 1.6 \%$ at zenith.

\begin{figure*}
\centering
\includegraphics[width=0.9\textwidth,clip]{ 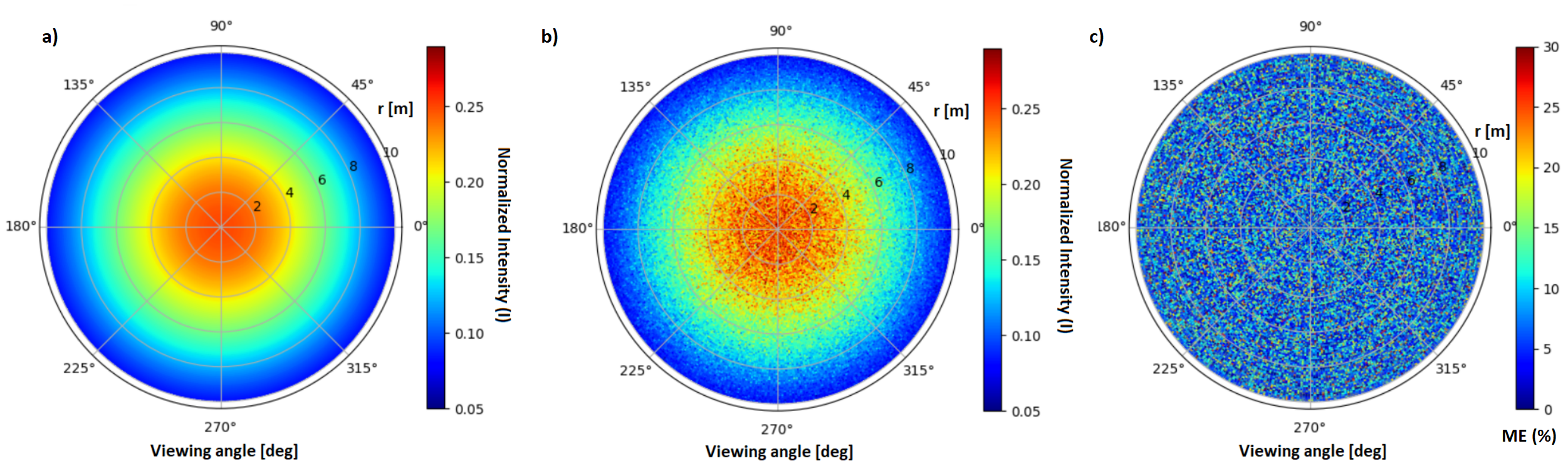}
\caption{Normalized intensity profile from the S/C viewed at the GS: a) Without scintillation, b) With scintillation. c) Absolute mean error ($\%$) between a) and b).}
\label{Scin}       
\end{figure*}

\medskip\noindent
These results are in accordance with the work of \cite{mom1}. The authors of that work experimentally obtained the DoP of a circularly polarized photon source from a S/C propagating along the atmosphere, obtaining a photon depolarization up to $4.0\%$. By comparison, we achieve similar results with discrepancies within the uncertainty range.

\medskip\noindent
\subsection{Atmospheric Turbulence Simulation}
\label{Turbulent effects}

For this analysis we shall consider three main effects: beam spreading, scintillation and beam wandering \cite{v8}. These effects are dependent on the relative size between the eddies of the turbulent layer and the beam width of the signal. For beam width sizes bigger than the eddies radius we consider that the beam spreading effect is dominant. If the size is of the same order of magnitude then scintillation dominates, this effect is mainly relevant for satellite downlinks. Beam wandering mostly occurs for uplinks, this can be justified due to an immediate entry of the signal in the atmospheric layer where the beam width is at its minimum being much more negligible compared to the eddies radius.

\medskip\noindent
To calculate beam spreading we have first defined our turbulent profile, based on three distinct models and compared their performance relative to each other, a low zenith model which performs better at lower intensity profiles for the Kolmogorov turbulence \cite{popi}; a high zenith model which considers higher orders of magnitude for turbulent perturbations in the Kolmogorov theory \cite{v8}; a more generic non-Kolmogorov model which by having $\alpha=\frac{11}{3}$ allows us to recover the Kolmogorov theory \cite{popi4}.

\medskip\noindent
The effective beam spreading effect from the added turbulence was calculated by using $W_{eff}=w(1+T_{A})$, where w is the Gaussian beam width at the GS and $T_{A}$ is the turbulent scale factor which deforms the overall beam \cite{C}.

\medskip\noindent
At the worst case scenario, considering a strong turbulence profile at the horizon we have obtained a maximum additional beam spreading loss of 0.006 dB from the turbulent effect. Hence, the turbulent beam spreading effect loses its relevance when compared to the geometric effect, where by choosing the appropriate geometric parameters the losses could vary up to $-10$ dB.

\medskip\noindent
Scintillation on the other hand is dominant in our simulation. To calculate it we have considered a statistical model for the turbulent behaviour. We consider that for weak turbulence the intensity statistics can be described with a PDTC with a log normal intensity profile:

\begin{equation}
\begin{aligned}
p_{I}(\eta)=\frac{1}{\sqrt{2\pi}\sigma_{I} I(r,L)}exp\left(-\frac{\left(ln(\frac{I(r,L)}{<I_{0}(r,L)>})+\frac{1}{2}\sigma_{I}^2\right)^2}{2\sigma^2} \right)
\end{aligned}
\label{sinn}
\end{equation}

\medskip\noindent
In Equation \ref{sinn}, $\sigma_{I}$ is the scintillation index based on low and high turbulence zenith models, I(r,L) is the irradiance profile, r is the radius from the center of the receiver and L is the optical path. $\langle I(r,L) \rangle$ can be defined by $\langle I(r,L)\rangle=A(\frac{w_{0}}{w_{I}})^2 exp(\frac{-2r}{w_{I}}^2)$ \cite{C}, where A is a normalization constant and $w_{0}$ is the Gaussian beam width. The results for the scintillation effect can be seen in Fig \ref{Scin}.

\medskip\noindent
We have compared the intensity profile with and without scintillation present as seen in Fig \ref{Scin} b) and a), respectively, and we have compared the absolute mean error (ME) between the two in c). A maximal deviation of 30$\%$ was computed for the intensity profile, meaning that the signal losses can reach up to 3 dB.

\medskip\noindent
On the other hand, beam wandering, is responsible on how the atmospheric turbulence creates time-dependent random lateral beam displacements. In order to study its effect we must also consider the PDTC model. There, we must compare the intrinsic pointing error influence with the beam wandering. Therefore, to express the S/C pointing error we use the Weibull distribution \cite{v3,v8}: $P=\frac{r}{\sigma_{r}^2}exp\left(-\left( \frac{r}{\sqrt{2}\sigma_{r}}\right)\right)$. Here, $\sigma_{r}$ is the standard deviation for the Weibull distribution and $r$ is the shift deflection distance from the GS center. Also, $\sigma_{r}$ can be described by \cite{v7}: $\sigma_{r}=\sqrt{\left( \theta_{p}L\right)^{2}+\sigma_{w}^2}$. Where $\theta_{p}$ is the pointing error of the S/C and $\sigma_{w}$ is the variance of the beam center due to turbulence. In weak turbulence theory considering a collimated beam in Kolmogorov turbulence with infinite outer scale, the $\sigma_{w}$ term can be defined by $\sigma_{w}=1.919 C_{n}^2 z^3 \left( 2w_{0}\right)^{-\frac{1}{3}}$. Here, $C_{n}^2$ is described by the Hufnagel-Valley model $H-V_{5/7}$, $z$ is the optical path in the atmosphere and $w_{0}$ is the beam waist when entering the atmosphere.

\medskip\noindent
 By considering the optical path as well as $\theta_{p} = 1.0 \mu m$ from the precision obtained in the Micius mission S/C \cite{v3}, we obtain the following: $\left( \theta_{p}L\right)^{2}=0.56 m^2 >> 10^{-3}m^2 = \sigma_{W}^2$. Ergo, the order of magnitude for the beam wandering effect being near to negligible. This effect, is to be expected considering its physical interpretation, as photons from the S/C are considered to propagate $97\%$ of the time in near vacuum. When finally the signal enters the upper atmospheric layers ($\sim$ 20 km) the size of the beam width is already considerable ($w_{0}=13.17\pm0.14$ m at high zenith angles) compared to the eddies mean radius \cite{nut6}.

\begin{figure*}
\centering
\includegraphics[width=0.9\textwidth,clip]{ 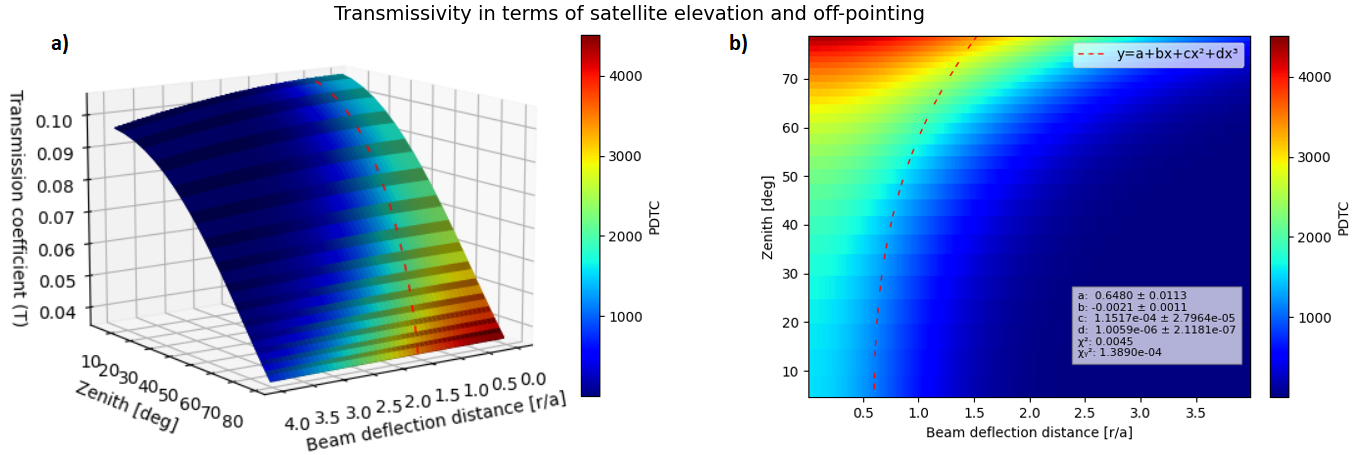}
\caption{a) Transmission coefficient in terms of S/C elevation, the PDTC and deflection distance $r/a$. b) Top-down view of a).}
\label{Trans}       
\end{figure*}

\medskip\noindent
\subsection{PDTE and off-pointing behaviour}
\label{PDTE and off-pointing behaviour}
  The objective in this section is to describe the transmission efficiency received at the GS for each satellite elevation and noise channel while measuring the mean beam deflection distance from the aperture center at the GS.
  Let us consider the PDTC statistical model which describes the transmission efficiency (T) received at GS by Equation \ref{tut1} \cite{nut7}:

 \begin{equation}
\begin{aligned}
T^2=T_{0}^2exp\left( -\left(\frac{r}{R_{1}}\right)^{-\lambda_{1}}\right)
\end{aligned}
\label{tut1}
\end{equation}

\medskip\noindent
$T_{0}$ is the maximal transmission coefficient for a given beam size ($W$), $\lambda_{1}$ and $R_{1}$ are the shape and scale
parameter respectively, which can be defined by \cite{nut7}:

\begin{equation}
\begin{aligned}
T_{0}^2=1-exp\left( -\left(\frac{a\sqrt{2}}{W}\right)^{2}\right)
\end{aligned}
\label{tut2}
\end{equation}

\medskip\noindent

\begin{equation}
\begin{aligned}
\lambda_{1}^2=8\frac{a^2}{W^2}\frac{exp\left(\frac{-4a^2}{W^2}\right)I_{1}\left( \frac{4a^2}{W^2}\right)}{1-exp\left(\frac{-4a^2}{W^2}\right)I_{0}\frac{4a^2}{W^2}}\\ \times\left[ln\left(\frac{2 T_{0}^2}{1-exp\left( -\frac{-4a^2}{W^2}\right)I_{0}\frac{4a^2}{W^2}}\right)\right]^{-1}
\end{aligned}
\label{tut3}
\end{equation}

\begin{equation}
\begin{aligned}
R_{1}=a\left[ln\left(\frac{2T_{0}^2}{1-exp(-\frac{-4a^2}{W^2})I_{0}\frac{4a^2}{W^2}}\right) \right]^{-\frac{1}{\lambda_{1}}}
\end{aligned}
\label{tut4}
\end{equation}

\medskip\noindent
The $a=\frac{D_{R}}{2}$ parameter is the radius of the receiver and r is its
beam-deflection distance. $I_0$ and $I_1$ are the modified Bessel functions of the first kind. Considering that the beam center is distributed accordingly to the profile described by the Weibull distribution as stated previously, we can combine it with Equations \ref{tut1}, \ref{tut2}, \ref{tut3} and \ref{tut4} to obtain the transmission coefficient accounting for the inner off pointing profile of the S/C. We assume that the beam fluctuates around the aperture center, allowing us to obtain the PDTC ($\mathcal{P}$) \cite{nut7,v7}:

\begin{equation}
\begin{aligned}
\mathcal{P}(T)=\frac{2R_{1}^2}{\sigma_r^2 \lambda_1 T}\left(2ln\left(\frac{T_{0}}{T}\right)\right)^{\frac{2}{\lambda_1}-1}\times\\
exp\left[-\frac{1}{2\sigma_r}R_{1}^2\left(2ln\left(\frac{T_{0}}{T}\right)^{\frac{2}{\lambda_1}}\right)\right]
\end{aligned}
\label{tut5}
\end{equation}

\medskip\noindent
We shall now use Equation \ref{tut5} with the S/C to GS optical paths. This allows us to obtain in Fig \ref{Trans} the transmission coefficient containing the geometrical and turbulence behaviour (T) in terms of the S/C zenith angle, the probability for beam deflection $\frac{r}{a}$ from the center of the receiver and the effective distance from the center of GS. The mean beam deflection from the center of the aperture receiver is calculated by renormalizing the PDTC to one for each S/C zenith angle and applying a polynomial regression of our data up to third order $(O(3))$. The latter was performed using LMFIT 1.0.2 Python package, obtaining a $\chi^2=9.8032\times10^{-04}$ and $\chi_{\gamma}=2.2798\times10^{-05}$ for the Chi-Square and reduced Chi-Square parameters, respectively. From the results, a mean off-pointing from the aperture center of r = 1.20 m was obtained for low zenith angles, increasing up to r = 3.02 m for high zenith angles.

\medskip\noindent
Comparing with the state of the art, the transmissivity coefficent (T) achieves similar results with the works of \cite{nut7} and \cite{v7}, as well as with the equivalent eliptical model for QKD from the work of \cite{v6}.

\medskip\noindent
\section{Results and Discussion}
\label{Results and Discussion}
\begin{figure}
\centering
\includegraphics[width=0.48\textwidth,clip]{ 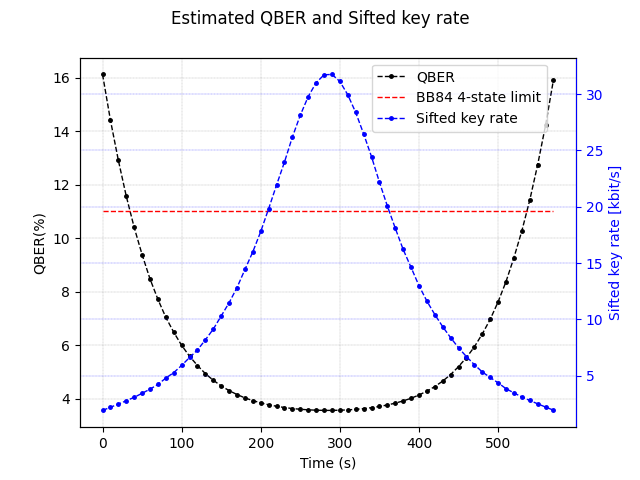}
\caption{Sifted key rate and QBER in terms of the S/C elevation for the QuantSat-PT mission considering all losses from Table \ref{fin}.}
\label{SifteBER}       
\end{figure}

For the BB84 protocol, to obtain the key rate, we use the overall loss rate presented in Table \ref{fin} for each S/C elevation. Table \ref{fin}, introduces to the expected optical budget for the Quantsat-PT mission at zenith, reaching to a total expected loss for the 4-state BB84 protocol of $[34.008, 37.099]\pm 0.400$ dB.  Hence, allowing to obtain Fig \ref{SifteBER}. Fig \ref{SifteBER} shows that the key rate for the mission reaches up to 32.1 kbit/s considering a $MPN=0.5$. The attenuation effects greatly decrease at high turbulent zenith angles the overall signal performance reaching down to 3 kbit/s at $\theta_{zen} = 80$ deg.

\medskip\noindent
For QBER, at low zenith angles up to $\theta_{zen}=60$ deg the quality of the 4-state BB84 protocol remains the same within a range of $ QBER\in[3.8,5.1] \%$. When in high turbulent environment the QBER can reach above the $11\%$. Hence, from that point, the protocol looses its validity, allowing for an eavesdropper to obtain relevant information without being noticed. Our active time for the mission shortens down to $\in[31,540]$ s as seen from Fig \ref{SifteBER}.

\begin{figure}
\centering
\includegraphics[width=0.48\textwidth,clip]{ 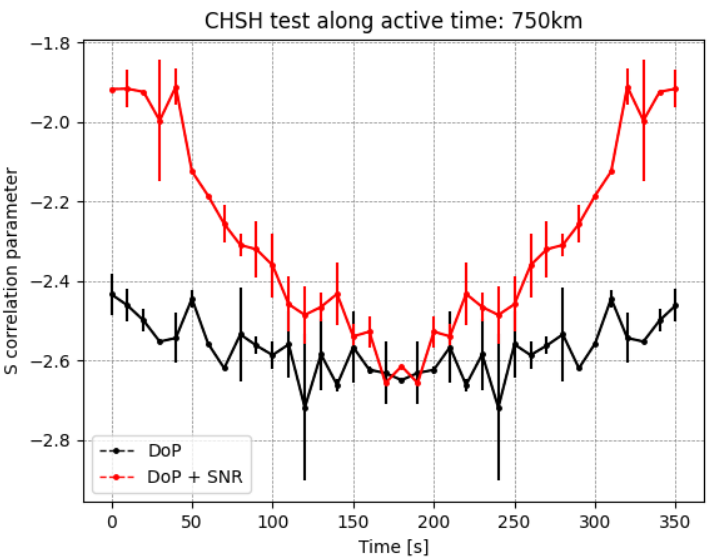}
\caption{Correlation coefficient $S$ during the active time of the mission. $\gamma_{DoP}$ and $\gamma_{SNR}$ parameters are present in order to study the vulnerability of the E91 protocol for our mission.}
\label{Correl}       
\end{figure}

\medskip\noindent
Similarly to what was done with the BB84 protocol, we evaluate the E91 protocol by performing the CHSH test. We shall only consider the signal to noise ratio (SNR) and the atmospheric DoP losses in our simulation as it is of the assumption that if a photon sent from S/C does not reach the GS it shall not be considered for the CHSH test validation. Instead, a background
photon will be considered in the expected time of arrival of the signal.

\medskip\noindent
In Fig \ref{Correl} we performed the CHSH test by using the $\gamma_{DoP}$ and $\gamma_{SNR}$ components in terms of the correlation coefficient ($S$) during the active time of the mission. Ideally $|S|=2\sqrt{2}$ which tells us that the basis are anti-correlated between Alice and Bob, showing the viability of the protocol for secure communications. Without any environmental or geometrical perturbations considered, if the correlation coefficient follows $|S|\leq 2$, it means that either the
received photons are not truly entangled or that a high degree of losses are present in our system \cite{bobi1}. As we cannot distinguish experimentally between both statements, we shall consider $|S|\leq 2$ as our limit to perform secure communications for the E91 protocol.

\medskip\noindent
The large difference in error bars in Fig \ref{Correl} is due to statistics, by the low number of singlets used in the simulation ($10^{4}$ photons). This is justified by the high simulation times required to obtain a more realistic scenario of $10^7$ photons for the satellite mission.

\medskip\noindent
As seen in Fig \ref{Correl}, with only $\gamma_{DoP}$ component present we obtain a valid S parameter of $\in[-2.67\pm0.23,-2.44\pm0.04]$. By introducing the SNR our CHSH test performance shifts where the Bell's test for the E91 protocol is only valid between $\in[46,321]$ s. 

\medskip\noindent
These results not only are in compliance with the QuantSat-PT mission requirements as they have been validated accordingly, with the work of \cite{v6}, \cite{v4} and \cite{v7}. Where considering similar S/C trajectory profiles and initial parameters we obtain results down to a 3 dB error. \cite{v6} considers an altitude of 500 km, with a $w_{0}=0.05$ m and $D_{R} = 1.0$ m. The author's work achieves for Cubesats a $QBER=3\%$ at low zenith angles going up to $QBER=14\%$ at zenith angles above $\theta_{zen}=75\%$. Our simulator reaches the same QBER profile being within the uncertainty range for the total losses.

\medskip\noindent
We also compared our simulator with the work of \cite{v4} which considers a 600 km orbit, with the same parameters as defined previously from the work of Carlo Liorni et al. The author's achieve a QBER below $11\%$ between zenith angles of $\theta_{zen}\in[0,70]$ deg. Once again, considering similar geometrical parameters our results reach the same order of magnitude as in their analysis.

\begin{table}
\centering
\caption{Optical parameters for Quantsat-PT mission at 750 km orbit zenith.}

\renewcommand{\arraystretch}{1.5} % Default
\begin{tabular}{ |p{3.4cm}|p{3.8cm}|p{1cm}| }

\hline
\textbf{Parameters} & \textbf{Description (at zenith)} & \textbf{Units}   \\
\hline
Signal ratio ($S_F$) & $79.8$ &  $\%$  \\
\hline
S/C mean off-pointing $(\frac{2r}{D_{R}})$  & $0.639$ & -  \\
\hline
Geometric loss  & $28.201\pm0.001$ &    \\
\cline{1-2}
Atmospheric loss  & $1.422^{+0.184}_{-0.374}$  &    \\
\cline{1-2}
Depolarization loss  & $0.284^{+0.142}_{-0.139} $&   \\
\cline{1-2}
Background SNR loss  & $1.988$ &    \\
\cline{1-2}

Beam spreading loss  & $0.003$ &  dB  \\
\cline{1-2}
Beam wandering loss  & $0.015$ &    \\
\cline{1-2}
Scintillation loss  & $[0.000, 3.091]$ &    \\
\cline{1-2}
Mean off-pointing loss  & $1.861$ &   \\
\cline{1-2}
Basis rotation shift loss & $0.265$ &   \\
\cline{1-2}
Wavefront aberration loss  & $0.619$ &   \\
\cline{1-2}
Total loss for BB84 protocol & $[34.008, 37.099]\pm 0.400$ &  \\
\hline

\end{tabular}
\label{fin} 
\end{table}

\medskip\noindent
\section{Conclusions}
\label{Conclusions}
This work contributes to solve one of the major problems in QKD space missions which is the precise consideration of atmospheric losses on the signal. The simulator was developed as part of the QuantSat-PT project, which aims to perform the first Portuguese QKD space mission on a 3U Cubesat. 

\medskip\noindent
In this work we have computed the sifted key as well as the QBER for the 4-state BB84 protocol which reach up to 32.1 kbit/s and $4\%$ at zenith respectively for a 750 km orbit. For the E91 protocol a similar analysis was performed, however this time, the CHSH test was studied. By creating a quantum circuit for the E91 protocol, we have obtained a correlation factor of $S \in[-2.63\pm0.02,-1.91\pm0.03]$ for the mission, considering the depolarization and SNR terms. Moreover, in-depth analysis for the turbulent behaviour was performed as well as the depolarization ratio on our signal with the study of the Stokes parameters. A statistical analysis was also proposed for our mission, which considered the mean off-pointing behaviour of the satellite in a turbulent environment. Furthermore, the photon propagation along the atmosphere was simulated with Monte Carlo allowing to obtain the atmospheric transmissivity considering the absorption and Rayleigh scattering effects.

\medskip\noindent
To improve even further the simulator, we must take into account the hardware behaviour of the optical payload. Thus, by performing the hardware in loop testing we could create a more realistic model for the signal's intensity profile. This would also allow for a more robust modelling for the optical and quantum efficiencies for each optical segment which lead to a more realistic QKD performance.

\medskip\noindent
In order to improve the accuracy of the night sky background behaviour, it is essential to perform a local set of measurements for the brightness of night sky in the OGS region. Allowing to realistically calibrate the simulator considering the local mean natural and artificial background noise. 

%%%%%%%%%%%%%%%%%%%%%%%%%%%%%%%%%%%%%%%%%%%%%%%%%%

\section*{Data Availability}
The data that support the findings of this study are available from the corresponding
authors upon reasonable request.

\section*{acknowledgments}
This work was supported by Instituto de Telecomunicações through QuantSat-PT project. The authors would like to thank the QuantSat-PT team and Prof. João Seixas for support, helpful discussions and article review.

\section*{Author Contributions}
V.G developed and ran the simulator according to the requirements for the QuantSat-PT mission. M.N conceived and supervised the project. All authors contributed to writing the manuscript.



\bibliography{apssamp}% Produces the bibliography via BibTeX. % if your bibtex file is called example.bib

%\begin{thebibliography}{99}
%\end{thebibliography}

%%%%%%%%%%%%%%%%%%%%%%%%%%%%%%%%%%%%%%%%%%%%%%%%%%

%%%%%%%%%%%%%%%%% APPENDICES %%%%%%%%%%%%%%%%%%%%%

\end{document}